\newif\ifshow
\showfalse 

\newif\ifinclude
\includetrue

\documentclass[acmsmall,screen,nonacm]{acmart}

\makeatletter
\def\@ACM@checkaffil{}
\makeatother

\usepackage[most]{tcolorbox}
\usepackage{subcaption}
\usepackage{wrapfig}
\usepackage{placeins}

\newtcolorbox[auto counter]{rqbox}[1][]{%
  enhanced, breakable, sharp corners,
  colback=black!3, colframe=black!75,
  boxrule=0.5pt, leftrule=3pt,
  left=6pt, right=6pt, top=6pt, bottom=6pt,
  #1
}

\usepackage{xcolor}

\definecolor{stateRevise}{HTML}{1C5AA8}
\definecolor{stateValidate}{HTML}{6842A0}
\definecolor{stateAssess}{HTML}{8A7000}
\newcommand{\stateRevise}{\textcolor{stateRevise}{\textbf{Revise}}}
\newcommand{\stateValidate}{\textcolor{stateValidate}{\textbf{Validate}}}
\newcommand{\stateAssess}{\textcolor{stateAssess}{\textbf{Assess}}}
\newcommand{\stateReviseCheck}{\textcolor{stateRevise}{\ensuremath{\checkmark}}}
\newcommand{\stateValidateCheck}{\textcolor{stateValidate}{\ensuremath{\checkmark}}}
\newcommand{\stateAssessCheck}{\textcolor{stateAssess}{\ensuremath{\checkmark}}}

\usepackage{tikz}
\newlength{\circledD}
\newcommand{\circled}[1]{%
  \tikz[baseline=(char.base)]{%
    \node[draw, circle, line width=0.5pt,
          inner sep=0pt, outer sep=0pt,
          minimum size=\circledD]
      (char) {\rmfamily\bfseries\footnotesize
              \makebox[0pt][c]{\vphantom{A}#1}};%
  }%
}

\newcommand{\repairstep}[1]{%
  \tikz[baseline=(char.base)]{%
    \node[draw=black, fill=white, rectangle, line width=0.5pt,
          inner sep=0pt, outer sep=0pt, minimum size=\circledD]
      (char) {\rmfamily\footnotesize\makebox[0pt][c]{\vphantom{A}#1}};%
  }%
}

\AtBeginDocument{%
  }

\authorsaddresses{}

\begin{document}

\title[Testuary: Agentic Test Generation with Zero-Shot Parallelism]{Testuary: Agentic Test Generation with Zero-Shot Parallelism}

\author{Kevin Kammler}
\affiliation{%
  \institution{Johannes Gutenberg University Mainz}
  \city{Mainz}
  }
\email{k.kammler@uni-mainz.de}
\author{Dominik Sobania}
\affiliation{%
  \institution{paluno -- The Ruhr Institute for Software Technology, University of Duisburg-Essen}
  \city{Essen}
  }
\email{dominik.sobania@uni-due.de}

\renewcommand{\shortauthors}{K. Kammler \& D. Sobania}

\begin{abstract}
Agentic approaches for automated test generation can reach high-quality results, but they improve a given test suite incrementally, so a weak starting point may lead to many iterations or insufficient tests. To address this, we present Testuary, which provides the agentic phases with an improved starting point. Testuary first generates multiple test suites through independent, parallel zero-shot requests, checks and repairs them deterministically, and combines them into a single suite with a novel structural merging approach. A runtime-weighted set-cover reduction removes redundant tests without losing covered lines, branches, or killed mutants. After that, two agentic phases target the remaining gaps in coverage and mutation score. We compare Testuary with established search and LLM-based test generation approaches. The non-agentic first phase alone already significantly outperforms the current state-of-the-art. After the agentic phases, Testuary achieves on average more than 99.9\% line and branch coverage as well as a mutation score of 98.6\%, significantly outperforming all compared baseline approaches. Testuary needs fewer model requests and its test suites are on average smaller and faster than those of the compared agentic approach.\end{abstract}


\ifinclude
\settopmatter{printfolios=true}

\fancypagestyle{nonacmplain}{%
  \fancyhf{}%
  \fancyfoot[C]{\thepage}%
  \renewcommand{\headrulewidth}{0pt}%
  \renewcommand{\footrulewidth}{0pt}%
}
\pagestyle{nonacmplain}
\fi

\maketitle

\ifinclude
\thispagestyle{nonacmplain} %
\fi

\section{Introduction}

High-quality test suites are essential in software engineering, as they help developers detect errors introduced by code changes right away and correct them before they propagate \cite{zhu1997software,yoo2012regression}. In many software projects, however, developers do not invest in adequate testing for several reasons. For example, short-term business decisions could lead to technical debt \cite{ernst2015measure} which may never be addressed afterwards \cite{freire2020surveying}, leaving tests under-prioritized. Furthermore, developers perceive writing tests as effortful and often de-prioritize writing them \cite{straubinger2023survey}. Yet, tests become urgently necessary when far-reaching changes are made, to ensure that the software's behavior remains unchanged. Such changes could be, e.g., an update of important libraries or a migration to the cloud.

To address this problem, various approaches have been proposed for automated test generation in the past. Prominent examples are techniques from search-based software testing~\cite{mcminn2011search} such as EvoSuite~\cite{fraser2011evosuite} or Pynguin~\cite{lukasczyk2022pynguin}, which iteratively search for test suites with high test coverage. Hybrid approaches also exist \cite{lemieux2023codamosa} that combine the search with large language models (LLMs) in order to escape plateaus or local optima. More recently, high test coverage scores have been achieved with agentic approaches like CoverUp~\cite{altmayer2025coverup}.
However, a drawback of such systems is that agentic loops improve a given test suite (e.g., an existing test suite or one obtained via a zero-shot request to an LLM) incrementally \cite{jain2025testforge}, which means that a weak starting point can require a large number of iterations for the improvement. We expect that combining several independently generated test suites can provide a stronger starting point for agentic test generation without first requiring a sequence of agentic revisions to a single suite.

Therefore, in this paper, we present \textsc{Testuary}\footnote{The name combines the words \textit{test} and \textit{estuary}. An estuary is defined as the area where a river's fresh water and the sea's salt water are merged~\cite{noaa-estuary}. Likewise, Testuary combines independently generated test suites into a single one.}, which generates a set of initial test suites via independent, parallel zero-shot requests and merges them into a single test suite enabled by a novel structural merging method. Merging combines complementary tests to increase the test coverage and mutation score (i.e., the fraction of introduced code mutations a suite successfully detects \cite{budd1980theoretical,jia2010analysis}). Subsequent reduction aims to reduce the test suite runtime while retaining these results. Testuary's agentic phases can then, if necessary, build on this test suite and concentrate on the remaining gaps to improve the metrics even further.

Testuary first sends $n$ parallel zero-shot requests to an LLM for a given source file under test. Especially for more complex code under test, a single generated test suite often leaves noticeable gaps in coverage and mutation score. Combining these test suites yields significantly higher scores than the average individual zero-shot suite in our experiments, and starting from this stronger suite can reduce the time and model use needed for agentic refinement. To ensure that the combined test suite remains practically usable, we perform a test suite reduction with a set-cover approach~\cite{yoo2012regression}, which aims to reduce test suite runtime while retaining the coverage and killed mutants of the merged test suite. To make this possible, we introduce in Testuary a novel structural merging mechanism designed to preserve complementary tests when combining the individual suites. The resulting suite gives the two optional agentic phases a head start, with the first phase targeting remaining coverage gaps and the second improving the mutation score. At the end of each agentic phase, we merge the agent's result with the test suite from the preceding phase and re-run the reduction, to preserve useful earlier tests and to keep the resulting test suite small and efficient.

We evaluated Testuary on a common test generation benchmark and compared it to several well-known baseline approaches, most notably CoverUp~\cite{altmayer2025coverup}. The results show that already the initial phase, which merges the parallel zero-shot-generated test suites, on average outperforms methods such as CoverUp as well as search-based baselines. For example, CoverUp achieves an average branch coverage of 87.6\% and a mutation score of 68.2\%, whereas Testuary already reaches 98.7\% and 82.9\%, respectively, after its initial phase. The subsequent agentic phases increase these scores to 99.9\% for branch coverage and 98.6\% for the mutation score. Moreover, Testuary achieves significantly higher scores per second of test suite runtime than CoverUp.

In summary, we make the following contributions:

\begin{itemize}
    \item Introduction of Testuary, a new test generation framework that combines test suites generated by parallel zero-shot requests giving the following agentic test generation process a head start and allowing the agentic phases to focus on closing the remaining gaps.
    \item A novel structural merging approach that enables merging and reduction of independently generated test suites, usually without any loss of the coverage and mutation score metrics.
    \item An evaluation of all three phases of Testuary on a representative benchmark set demonstrating that already Testuary's non-agentic phase is able to outperform state-of-the-art methods like CoverUp in average line and branch coverage as well as mutation score.
\end{itemize}

\section{Related Work}\label{sec:related_work}

Automated test generation has been studied in software engineering for decades. In their survey, Anand et al. \cite{anand2013orchestrated} identified and described the main traditional approaches in this area, which we briefly outline in the following.

One of the earliest approaches for test generation is \textit{symbolic execution} \cite{king1975new,clarke1976system,king1976symbolic}, which dates back to the 1970s. The basic idea is to execute a program with symbolic rather than concrete input values, thereby systematically exploring program paths and deriving logical constraints. Another established line of research is \textit{model-based testing} \cite{utting2007practical,tretmans2008model,veanes2008model}, in which a model of the expected behavior of the software under test is used to generate test cases. \textit{Combinatorial testing} \cite{nie2011survey,cohen1997aetg} addresses the interaction of input parameters by selecting combinations that cover input interactions up to a pre-defined size. A key empirical finding is that relatively few combinations are often already sufficient to reveal a large fraction of faults \cite{kuhn2004software}, which makes combinatorial testing attractive in practice due to its comparatively low execution cost. Methods based on \textit{adaptive random testing} \cite{chen2005adaptive,tappenden2009novel,liu2011adaptive} aim to distribute non-failing test cases as evenly as possible across the input domain in order to avoid clustering and thereby increase the probability of detecting failures.

Another notable approach is \textit{search-based software testing (SBST)} \cite{mcminn2011search,harman2015achievements}, which became especially prominent throughout the 2000s and into the early 2020s. Here, search techniques such as evolutionary algorithms or local search strategies are used to find test suites with the desired properties. The search can target objectives such as high coverage or mutation scores. Test coverage \cite{ivankovic2019code} characterizes how thoroughly the source code has been executed by the tests. Common variants include line coverage, which measures the share of reached source code lines, and branch coverage, which measures the share of control-flow branch outcomes taken by the tests. The mutation score, in turn, is obtained from mutation testing \cite{budd1980theoretical,jia2010analysis}, where small code changes are applied to produce program variants called mutants. The score itself summarizes the share of mutants detected (i.e. \textit{killed}) by the tests.

Well-known SBST approaches are, for example, EvoSuite \cite{fraser2011evosuite} for the generation of JUnit tests in Java and, more recently, Pynguin \cite{lukasczyk2022pynguin} for generating Python tests. What these SBST approaches have in common with the other previously mentioned traditional methods is that they achieve no semantic understanding of the program under test. In SBST, this limitation can cause the search to stagnate on a coverage plateau, which Lemieux et al. \cite{lemieux2023codamosa} address with CodaMOSA by leveraging the semantic understanding capabilities of LLMs to generate more suitable inputs for the program under test and to overcome such plateaus.

Beyond SBST, LLMs have also been employed for test generation \cite{bareiss2022code, wang2024hits}. A notable example is MuTAP \cite{dakhel2024effective}, which uses an LLM in a loop to increase the mutation score. However, this approach targets individual functions and smaller tasks rather than complex project-level source files.

Particularly relevant to this paper, however, are recent agentic AI approaches that have achieved high-quality results. TestForge\footnote{An experimental comparison with TestForge was not possible, as the replication package linked in the paper was unavailable at the time of writing.}~\cite{jain2025testforge} iteratively refines a zero-shot generated test suite based on coverage feedback, while allowing the agent to invoke tools during execution, such as searching and viewing code, writing files, and editing code lines. CoverUp~\cite{altmayer2025coverup} combines code segmentation, static analysis (via tools), and integration checks for LLM-based test generation, and outperforms approaches such as MuTAP in terms of coverage.

\section{Methodology}
\label{sec:methodology}
\label{sec:method}

In this section, we describe how Testuary generates high-quality test suites for a given source file. We start by giving an overview of the whole pipeline in Sect.~\ref{sec:pipeline}. 
The subsequent sections then describe the pipeline in more detail, covering the parallel zero-shot generation and repair in Sect.~\ref{sec:phase1}, the two agentic phases in Sect.~\ref{sec:agent-loops}, and the consolidation and reduction in Sect.~\ref{sec:phase-end-consolidation}.

\subsection{Overview}
\label{sec:pipeline}

\begin{figure}[H]
  \centering
  \includegraphics[width=\textwidth]{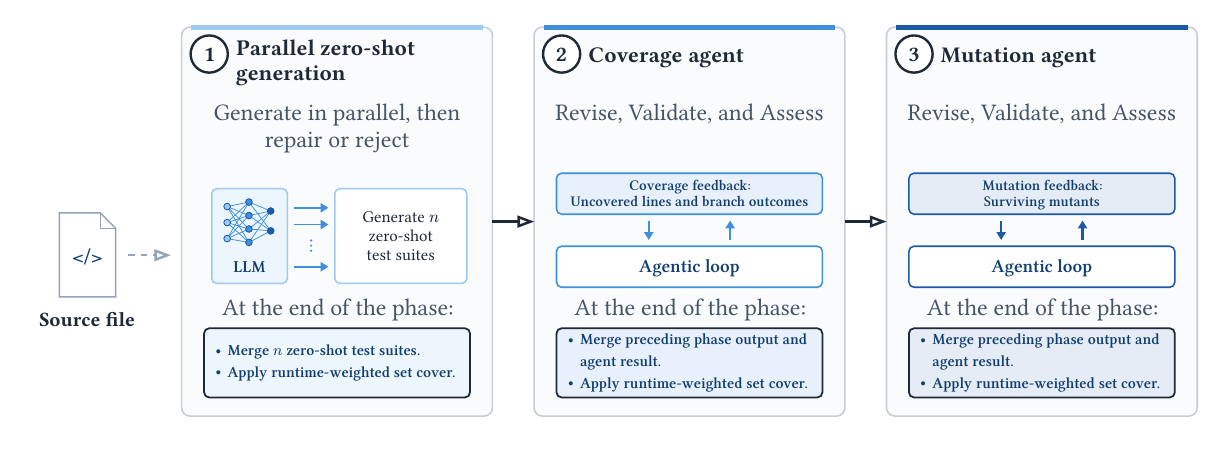}
  \Description{Three connected panels run from left to right, each with generation or refinement above merging and reduction.}
  \caption{Testuary's three-phase pipeline.
  Parallel zero-shot generation is followed by coverage-guided and mutation-guided refinement, with consolidation after each phase.}
  \label{fig:testuary-pipeline}
\end{figure}

Testuary's pipeline begins with parallel zero-shot generation and deterministic repair, then continues with coverage-guided and mutation-guided agentic refinement.
Figure~\ref{fig:testuary-pipeline} summarizes these phases and the consolidation performed at the end of each phase. The circled numbers identify the three phases, and the arrows show the test suite passed from one phase to the next.
The upper part of each box shows the phase-specific methodology, meaning either the zero-shot generation or an agentic loop.
The lower part of each box represents phase-end test suite consolidation, which first merges the suites available at the end of a phase and then removes redundant tests from the merged suite.
The merge inputs differ between the initial and agentic phases, while reduction follows the same procedure in all three phases.

Phase~\circled{1} sends $n$ independent zero-shot requests to an LLM.
Each request asks the LLM to generate a complete test suite directly for the source file under test without receiving example tests or another request's response.
These requests can be processed in parallel because none depends on another response.
Then, Testuary applies test suite repair, a sequence of deterministic repair and acceptance checks, to every returned suite.
Only suites that pass these checks proceed to consolidation.
During consolidation, Testuary combines tests from the independently generated suites, which can cover different lines and branch outcomes or kill different mutants.
It then removes redundant tests through runtime-weighted set-cover reduction~\cite{yoo2012regression}.
For this reduction, Testuary measures the merged suite and records the lines and branch outcomes covered by each test, the mutants it kills, and its runtime.
Using these measurements, it selects a subset with the minimum total recorded runtime that retains the merged suite's covered lines, branch outcomes, and killed mutants.
The resulting suite becomes the starting suite for the following coverage agent.

In Phase~\circled{2}, the coverage agent starts from an editable copy of the initial-phase result.
It uses coverage feedback to target lines and branch outcomes that the tests have not reached.
When the agentic coverage-guided refinement ends, Testuary applies the consolidation procedure used after the initial phase.
However, unlike before, the merge here combines the agent's result with the previous Phase~\circled{1} result.
Retaining the earlier result allows useful tests to be preserved if the coverage-guided revisions altered or removed them.
The test suite at the end of this phase serves as the starting point for the following mutation agent.

Phase~\circled{3} uses an editable copy of the coverage-phase result and applies the same agentic loop with mutation feedback, 
which guides the agent to add or adjust tests and assertions.
When mutation-guided refinement ends, Testuary applies consolidation again, this time merging the agent's result with the Phase~\circled{2} result.
The resulting suite becomes Testuary's final test suite.

Note that we divided coverage and mutation score improvement into two agentic phases because preliminary analyses suggested that this helps the agents focus on one kind of feedback and its corresponding tools.

\subsection{Phase 1: Parallel Zero-Shot Test Suite Generation with Deterministic Repair}
\label{sec:phase1}

Phase~\circled{1} requests $n$ independently generated zero-shot test suites from the configured LLM for each source file under test, using the same zero-shot prompt.
Since the requests do not depend on one another, they can be processed in parallel.
When several requests begin with the same text, a prompt cache allows the inference provider to reuse computation for this shared prefix instead of processing it again \cite{kwon2023efficient,deepseek-cache}.
Testuary therefore completes one request first to populate the cache and then submits the remaining requests concurrently.

Each generated suite undergoes test suite repair before it can contribute tests to Phase~\circled{1}.
For each of the $n$ requested suites, Testuary only allows a limited number of generation attempts.
If a response contains no test code or repair cannot produce an acceptable suite, Testuary asks the model to generate a replacement, provided the defined limit has not been reached.

\begin{figure}[!tbp]
  \centering
  \includegraphics[width=\textwidth]{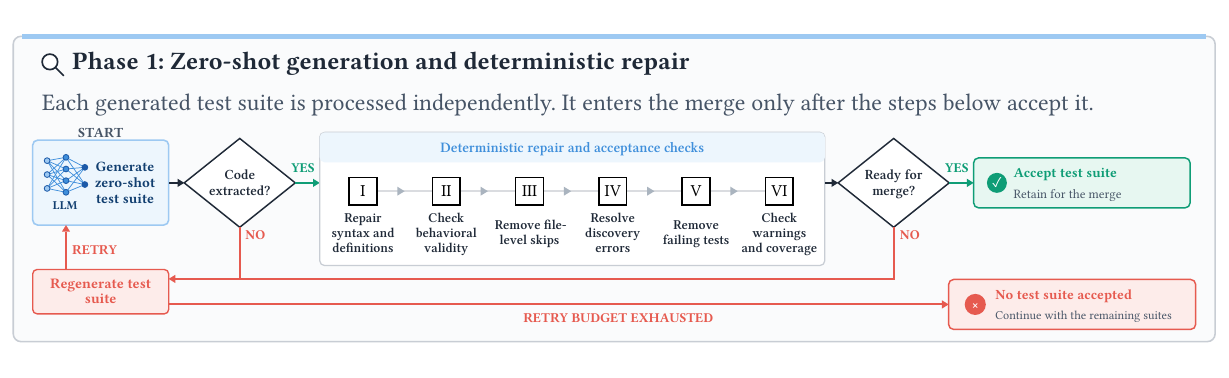}
  \Description{Flowchart linking test generation, six ordered repair checks, and acceptance or a replacement request if attempts remain.}
  \caption{Phase~\protect\circled{1} zero-shot generation and deterministic repair.
  The six checks make no model requests.}
  \label{fig:phase1-repair}
\end{figure}

Figure~\ref{fig:phase1-repair} shows how a generated suite is repaired and when Testuary requests a replacement.
The squares \repairstep{I} to \repairstep{VI} inside the test suite repair box in Fig.~\ref{fig:phase1-repair} show six steps applied in order.
Each step changes the test file only when it detects the corresponding problem.
If the problem cannot be resolved within a step, processing of the current generated version stops and follows the retry path shown in the figure.

Step~\repairstep{I} checks first whether the generated file can be parsed.
If not, Testuary can apply simple local repairs such as adding a missing body to an empty block or removing a line identified as invalid by the parser.
The generated version is discarded if these repairs cannot produce a parseable file.
Further, this step handles and removes duplicate definitions in the same scope.

Step~\repairstep{II} prevents generated tests from increasing the mutation score merely by detecting that the source code has changed.
A test could for example use and compare to the code under test directly to circumvent common testing practices.
Then, if a mutant changes the code under test, it causes the test to fail and the mutant to be counted as killed even though the test does not check whether the program behaves as expected.
In such cases, Testuary removes the offending test.
If it occurs in module-level code or a shared helper, it can affect several tests and cannot be removed together with one individual test, so Testuary discards the whole generated suite.

Step~\repairstep{III} removes instructions in the generated suite that can make the test framework skip the complete file.
In Python, a module-level call to \texttt{pytest.importorskip("optional\_package")} causes \texttt{pytest} to skip the file when the package is unavailable \cite{pytest}.
Such a guard can be useful for optional dependencies, but it also skips tests that do not need that package.
Testuary removes the guard and handles any resulting collection errors or failing tests in the following repair steps.

Step~\repairstep{IV} handles errors that occur before the test framework can discover the individual tests.
In Python, \texttt{pytest} executes imports and other file-level statements while discovering the tests.
An import of a name that does not exist in the code under test can therefore stop discovery before any individual test result is available.
Testuary first removes file-level imports that cannot be resolved.
If discovery still fails, it uses the error location reported by \texttt{pytest} to remove the responsible file-level statement before trying again.
A test that depends on the removed statement is then reported as failing and handled in Step~\repairstep{V}.

Step~\repairstep{V} handles failures reported for individual tests.
Testuary removes the failing tests and runs the remaining tests again because removal may lead to further failures.
This cleanup process stops after a configured number of rounds, because a suite that still fails can be replaced by a new LLM response.
On the last allowed generation attempt, no replacement is possible, so Testuary continues removing reported failing tests as long as each round makes progress.
If no passing subset can be produced, the generated suite is discarded.

Step~\repairstep{VI} verifies that the execution time stays within the configured limit and checks for warnings that indicate incomplete test execution.
When such a warning can be attributed to a test, Testuary removes that test and checks the remaining suite again.
Testuary also measures coverage and keeps the suite only if it reaches at least one executable statement in the source file under test.
This prevents an empty, entirely skipped, or ineffective suite from being accepted.

Test suite repair handles one generated suite at a time.
Accepted suites then enter the phase-end test suite consolidation we describe in Sect.~\ref{sec:phase-end-consolidation}.
There, structural merging resolves conflicts introduced by combining the suites, and set-cover reduction removes redundant tests.

\subsection{Phases 2 and 3: Agentic Loops for Coverage and Mutation Score Improvement}
\label{sec:agent-loops}

The separate coverage and mutation phases are the agentic part of Testuary (see Phases \circled{2} and \circled{3} in Fig.~\ref{fig:testuary-pipeline}).
In both phases, the agent starts from the suite produced by the previous phase. To improve the given test suite, the agent edits the tests, executes the updated suite, and measures the phase's metric (coverage or mutation metrics) while ensuring that no errors are introduced.

\begin{wrapfigure}{r}{0.37\textwidth}
  \centering
  \includegraphics[width=\linewidth]{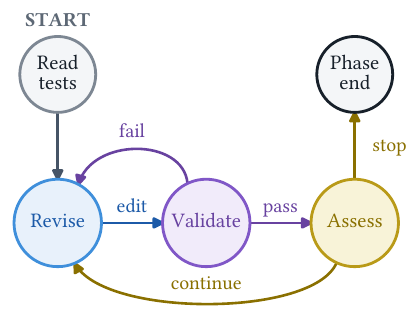}
  \Description{Read tests leads into a loop of Revise, Validate, and Assess. Failure and continuation arrows return to Revise, while a stop arrow leads to Phase end.}
  \caption{Testuary's shared agentic loop.
  \stateRevise{} edits, \stateValidate{} executes, and \stateAssess{} measures the suite.
  Failed executions or insufficient results are returned to \stateRevise{}. %
  }
  \label{fig:agent-state-graph}
\end{wrapfigure}

Before an agent starts its work, Testuary establishes the suite's score from the previous phase as the baseline.
So in Phase~\circled{2}, the line and branch coverage values recorded in Phase~\circled{1} are used.
In Phase~\circled{3}, the mutation score recorded for the Phase~\circled{2} output and the source file under test are used.
If a test suite already satisfies the defined goal of the current agentic phase, Testuary skips the agentic loop.
Otherwise, the agent initially reads the provided test suite and starts its agentic loop which is illustrated in Fig.~\ref{fig:agent-state-graph}.
In Testuary, agentic loops are divided into the states \stateRevise, \stateValidate, and \stateAssess{} so that each model request can invoke only tools that are meaningful under the current conditions. This has the benefit that it prevents the agent, for example, from measuring an untested or failing edit which saves resources. In Table~\ref{tab:agent-stage-tools}, we provide an overview of the tools allowed in each agentic state.

\begin{table}[!htbp]
\centering
\setlength{\tabcolsep}{5pt}
\renewcommand{\arraystretch}{1.12}
\caption{Tool access by recurring agent state. Checkmarks (\checkmark) indicate availability.}
\label{tab:agent-stage-tools}
\begin{tabular*}{\textwidth}{@{\extracolsep{\fill}}llp{0.31\textwidth}ccc@{\hspace{4pt}}}
\toprule
Tool & Agent & Purpose & \multicolumn{3}{c}{Agent state} \\
\cmidrule(lr){4-6}
 & & & \stateRevise & \stateValidate & \stateAssess \\
\midrule
\texttt{read\_source} & Both & Inspect the source file under test & \stateReviseCheck & \stateValidateCheck & \stateAssessCheck \\
\texttt{read\_tests} & Both & Inspect the editable working test suite & \stateReviseCheck & \stateValidateCheck & \stateAssessCheck \\
\texttt{edit\_tests} & Both & Edit the working test suite & \stateReviseCheck & \stateValidateCheck &  \\
\texttt{run\_tests} & Both & Run the test suite and report failures &  & \stateValidateCheck &  \\
\texttt{measure\_coverage} & Coverage & Measure line and branch coverage and report gaps &  &  & \stateAssessCheck \\
\texttt{measure\_mutation} & Mutation & Measure mutation score and report surviving mutants &  &  & \stateAssessCheck \\
\bottomrule
\end{tabular*}
\end{table}

In \stateRevise, the agent can inspect the source file under test as well as the working test suite and can edit the tests.
After an edit, the previous execution and measurement no longer describe the current suite, so the control flow enters \stateValidate.
Reading and editing remain available in \stateValidate, and the agent can run the tests, but it cannot measure the edited suite until that suite passes.
A failing execution returns the control flow to \stateRevise, since another edit is required.
An executable and passing test suite enters \stateAssess, where the coverage or mutation measurement is enabled.
Reading remains available because it does not modify the working suite, while editing and test execution are not.
After \stateAssess, Testuary either ends the loop if a stopping condition applies or returns to \stateRevise{} because another edit is required for improvement.
Both agentic phases use configurable goals and stopping rules.
The loop ends when its goal is reached, when its model-request budget is exhausted, or after a configured number of consecutive measurements without sufficient improvement.
It also ends if an orchestration failure prevents another complete measurement.

The agent can edit only the working test suite, while the source file under test remains read-only.
Before applying an edit, Testuary checks that the suite parses and that there are no duplicate definitions, as described for repair Step~\repairstep{I} (see Sect.~\ref{sec:phase1}).
It also applies the behavioral check described for Step~\repairstep{II}, to ensure a test suite following common testing practices and allowing a correct measurement of the mutation score.
If a check fails, the edit is not written, and the tool response explains the problem so that the agent can adjust it in the next iteration.

For the whole agentic loop, we always keep the so far best passing revision of the test suite, to ensure that later changes do not reduce the test suite's score for the current phase objective.
A failing, timed-out, score-lowering, or unmeasured revision cannot replace it.
When the loop ends, the best found revision is used as the agent result for the phase-end consolidation, together with the unchanged suite from the previous phase. We will describe this process in detail in the following Sect.~\ref{sec:phase-end-consolidation}.

To keep the context size manageable during long agentic loops, we employ compaction to allow the agent to focus on the relevant information. 
During compaction, we retain the system prompt and recent messages but use placeholders for outdated file-read results (from tool calls) to keep the conversation history short and consistent.
However, compaction changes earlier parts of the context and can therefore prevent their reuse through prompt caching.
To limit this effect, we run compaction only when the agent's state changes, because state changes also modify the available tools, which in turn substantially influences the context. Between state changes, Testuary appends new messages to the conversation, keeping earlier content unchanged so prompt-cache hits reduce latency and cost \cite{deepseek-cache}.

In Testuary's implementation, we use coverage.py \cite{coveragepy} to measure line and branch coverage. To obtain the mutation score, we employ Cosmic Ray \cite{cosmic-ray}. In addition to the scores, we provide the agent also either the uncovered lines and branch outcomes or the surviving mutants as feedback.

\subsection{Phase-End Test Suite Consolidation}
\label{sec:phase-end-consolidation}

Testuary performs phase-end test suite consolidation after each of the three phases.
The procedure returns either a test suite that can be passed to the next phase or the final test suite.
After Phase~\circled{1}, it combines the accepted independent zero-shot suites so that complementary tests from different requests can contribute to a single test suite. After each of the agentic phases, it combines the unchanged suite from the preceding phase with the agent's result to preserve useful earlier tests. For example, the phase-end test suite consolidation in Phase~\circled{3} (mutation agent) receives the output suite from Phase~\circled{2} and the result of its own agentic loop. 
In the following, we name these test suites \textit{pre-consolidation suites}.
As described in Sect.~\ref{sec:agent-loops}, the kept agent result cannot score below the starting suite on the current phase's objective.
But this applies only to that objective, not to every individual covered line, branch outcome, or killed mutant.
A coverage-guided revision could therefore potentially lose a test that kills a particular mutant, while a mutation-guided revision could lose a test that covers a particular line or branch outcome.
Therefore, merging with the test suite from the preceding phase allows Testuary to retain those earlier tests alongside the agent's improvements to prevent regressions.

During consolidation, we also perform a test suite reduction. By using a set-cover approach~\cite{yoo2012regression}, we select a subset of the merged suite that keeps the covered lines, branch outcomes, and killed mutants while total recorded execution time is optimized.

In the following, we describe these two parts, the merging and the set-cover reduction, of the phase-end test suite consolidation in detail.

\subsubsection{Structural Merge}
\label{sec:structural-merge}

\begin{figure}[!htbp]
  \centering
  \includegraphics[width=\textwidth]{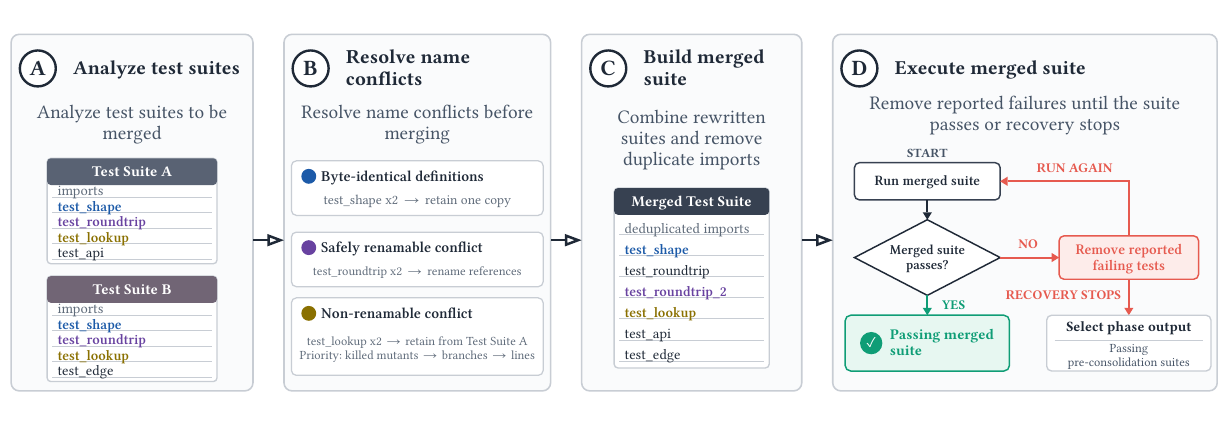}
  \Description{Four panels illustrate test suite analysis, name-conflict handling, merging, and execution with failing-test removal.}
  \caption{Structural merge during phase-end test suite consolidation.
  Steps~\protect\circled{A} to~\protect\circled{D} build and execute the merged suite.
  If no merged suite passes, a passing pre-consolidation suite becomes the phase output without reduction.}
  \label{fig:phase-end-consolidation}
\end{figure}

The consolidation process begins with the structural merge of the pre-consolidation suites.
Figure~\ref{fig:phase-end-consolidation} shows this in Steps~\circled{A} to~\circled{D}.

In \circled{A}, we analyze each test suite supplied to consolidation to identify its imports, definitions, tests, and name uses.
These suites have already been accepted separately, so structural merging addresses conflicts and failures introduced by combining them rather than repeating test suite repair.

Using this analysis, \circled{B} resolves name conflicts before the test suites are combined.
Shareable identical definitions are kept once.
When distinct definitions share a name, we keep the name for one of them, rename the others, and update the references that can be resolved.
If conflicting definitions cannot be renamed safely, only one definition and the tests that depend on it can remain.
Testuary favors the definition whose dependent tests kill more mutants, then cover more branch outcomes, and then more lines.
Following that, in Step~\circled{C}, we combine the rewritten test suites and remove duplicate imports.

In \circled{D}, we execute the test suite to check that also the merged suite passes. 
Tests that passed separately may fail after merging because test discovery order, shared runtime states, imports, or name bindings can change.
Therefore, we remove the reported failing tests and execute the suite again whenever at least one such test was removed.
Once a nonempty merged suite passes, we measure its mutation score, branch coverage, and line coverage, and record each test's coverage, killed mutants, and runtime for the following reduction process.
If it is not possible to merge a correct test suite due to unresolvable errors, reduction is skipped and the best pre-consolidation suite is used instead.

\subsubsection{Runtime-Weighted Set-Cover Reduction}
\label{sec:set-cover-consolidation}

\begin{figure}[!htbp]
  \centering
  \includegraphics[width=\textwidth]{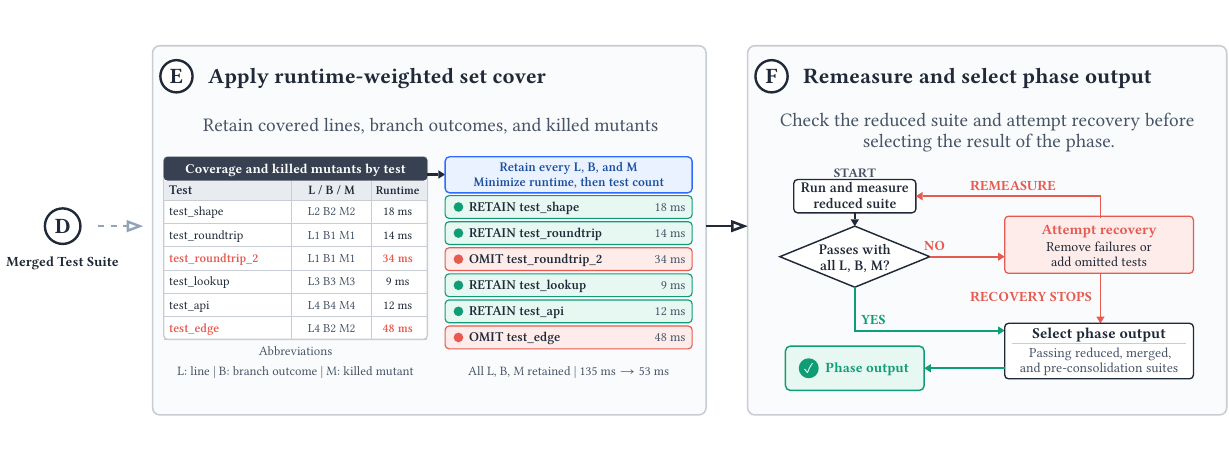}
  \Description{The left panel illustrates test selection from per-test measurements. The right panel shows reduced-suite checks, recovery, and phase-output selection.}
  \caption{Runtime-weighted set-cover reduction during phase-end test suite consolidation.
  Step~\protect\circled{E} minimizes runtime while retaining the merged suite's covered lines (\texttt{L}), branch outcomes (\texttt{B}), and killed mutants (\texttt{M}).
  Step~\protect\circled{F} executes and remeasures the reduced suite, then selects the phase output.}
  \label{fig:set-cover-reduction}
\end{figure}

With reduction, we remove redundant tests to improve the test suite runtime while keeping the covered lines, branch outcomes, and killed mutants recorded for that suite.
Figure~\ref{fig:set-cover-reduction} shows the steps involved in this process.

In \circled{E}, we associate each test in the merged suite with the lines and branch outcomes it covers, the mutants it kills, and its recorded execution time.
We model this selection as an optimization problem in which each test is either kept or omitted.
Each recorded line and branch outcome must be covered by at least one kept test, and each recorded killed mutant must be killed by at least one retained test.
The primary objective minimizes the total recorded execution time of the selected tests, while a secondary objective minimizes the number of tests among equally fast solutions.
To implement this method in Testuary, we use the COIN-OR Branch-and-Cut (CBC) solver \cite{forrest2005cbc} through the PuLP package \cite{mitchell2011pulp}.

In \circled{F}, we execute the reduced suite to confirm it passes.
This is not guaranteed, because removing tests can change how the remaining tests behave.
We additionally check that the reduced suite still covers every line and branch outcome covered by the merged suite and kills all mutants killed by the merged suite.
If either condition fails, we try a repair.
Failing tests are dropped, and tests that had been omitted are added back to recover lost lines, branch outcomes, or mutants.
The suite is remeasured after each repair.

At the end, the reduced suite usually becomes the result of the phase.
However, Testuary keeps the merged suite and the pre-consolidation suites as fallbacks so that consolidation does not discard a better result.
It selects the passing suite with the highest mutation score, resolving ties by higher branch coverage and then higher line coverage.
For equal scores, we favor the reduced suite, followed by the fallback with the lowest test suite runtime.
The chosen suite serves as the input for the next phase, or, after Phase~\circled{3}, the final test suite.

\section{Experimental Design}
\label{sec:experiments}
\label{sec:design}

We evaluate Testuary on file-level test generation and compare its final output
with traditional and LLM-based test generators. We also examine the contribution
of its three phases, model usage, and the size and execution time of the resulting
test suites.

\subsection{Research Questions}
\label{sec:rqs}

\begingroup
\setlength{\emergencystretch}{1em}
\begin{description}
  \item[\textbf{RQ1: Parallelization of zero-shot test generation.}] How do
  line coverage, branch coverage, and mutation score change for an increasing number of
  independently generated and combined zero-shot test suites?
  \item[\textbf{RQ2: Comparison with state-of-the-art methods.}] How does
  Testuary's output compare with established traditional
  and LLM-based test generators in line coverage, branch coverage, and
  mutation score?
  \item[\textbf{RQ3: Effectiveness across Testuary's phases.}] What line coverage, branch coverage, and mutation score does Testuary achieve after each phase, and how much does each phase contribute to these scores?
  \item[\textbf{RQ4: Model usage and test suite efficiency.}] How many model
  requests and tokens does each LLM-based method require, how does
  test suite effectiveness relate to test suite runtime, and what are the sizes and execution times
  of the delivered test suites?
\end{description}
\endgroup

\subsection{Experimental Setup} 
\label{sec:setup}
We evaluate Testuary, Zero-Shot Test Generation, Pynguin~\cite{lukasczyk2022pynguin}, CodaMOSA~\cite{lemieux2023codamosa}, and CoverUp~\cite{altmayer2025coverup} on all 45 files of the Python 3.10 subset of TestGenEval~\cite{jain2025testgeneval}.
These files represent test generation problems of widely varying sizes, ranging from 170 up to 5{,}701 lines of code.
The accumulated cyclomatic complexity~\cite{mccabe1976complexity} across functions within each file ranges from 39 to 803.
The use of this subset additionally enables us to include Pynguin and CodaMOSA in the comparison, since both tools require Python 3.10.
For each file, all methods are evaluated in the same project environment using the measurement procedure described below.

To ensure a fair comparison, all LLM-based methods (Zero-Shot, CodaMOSA, CoverUp, and Testuary) use the same model, DeepSeek V4 Pro accessed through the official API\footnote{\url{https://api-docs.deepseek.com/}}, with thinking enabled at the provider's default high reasoning effort.
We chose this model because it performs very well in current benchmarks and, as an open weights model, supports the reproducibility of our results.
To make CodaMOSA and CoverUp run reliably in this setup, we had to slightly adjust them.
CodaMOSA now supports the chat endpoint of the DeepSeek API, both methods now support retries on connection errors, and CoverUp gets the error message when JSON tool call arguments are malformed, so it can correct the call and keep working instead of aborting.
These changes help the methods continue instead of aborting after technical failures.

While the model is identical for all LLM-based methods, we cannot directly equalize their budgets because the methods use different stopping criteria.
Pynguin and CodaMOSA are limited by search time, CoverUp by the number of prompt attempts per code section, and Testuary by the number of model responses.
We therefore cap each method in the unit it is designed for.
For Pynguin and CodaMOSA we set a generous budget of one hour of search time per file.
CoverUp uses its default limit of prompt attempts per code section~\cite{altmayer2025coverup}.
Testuary requests 25 test suites in total in Phase~\circled{1} and allows per suite one initial generation attempt plus up to three replacement attempts, so at most 100 model responses.
The two agentic phases~\circled{2} and~\circled{3} allow 200 model-call attempts each, which caps a full run at a maximum of 500 model responses.
To still make the actual resource consumption comparable, we additionally record the number of model responses and token usage for all LLM-based methods.
We discuss these results later in the paper (see Sect.~\ref{sec:rq4}).
For CodaMOSA we added logging of the usage data returned by the API to support this.
Zero-Shot reports averages over the 25 individual suites accepted after repair in Phase~\circled{1}, providing a reference without merging.
Since both the model's responses and search-based test generation are nondeterministic, we run each method three times.

In our experiments we also measure the runtime of the test suites.
Each suite is run 30 times with pytest 8.1.1 under Python 3.10 after one unevaluated warm-up run.
We report the median of the wall-clock time, and the per-run timeout is set to 600 seconds.
We measure line and branch coverage of the tested source file with coverage.py~7.6.8~\cite{coveragepy}, and the mutation score with Cosmic~Ray~8.4.1~\cite{cosmic-ray} using its full operator catalog.
We follow TestGenEval's scoring convention for Cosmic Ray~\cite{jain2025testgeneval} regarding, e.g., failing tests and timeouts.
Mutants without a completed result are not included in the evaluation.
For each mutant the timeout is ten times the runtime of the test suite on the unchanged code, with a minimum of 30 seconds and a maximum of 900 seconds.
The total measurement is bounded by 7{,}200 seconds.
All measurements were run on AMD EPYC 7713 processors.
For runtime measurements, we allocated 4 CPU cores per job and 18 for coverage and mutation measurements.
Within this allocation, Cosmic Ray was configured to use 15 parallel workers.

\section{Results and Discussion}\label{sec:results}

In this section, we present and discuss the achieved results of Testuary's evaluation and answer the research questions.  

\subsection{RQ1: Parallelization of Zero-Shot Test Generation}

In Testuary's Phase \circled{1}, we combine independently generated zero-shot test suites to give the following agentic phases an improved starting point. Consequently, we analyze how an increasing number of independently generated zero-shot test suites affects the quality of the merged test suites. To answer this question, we calculate the  scores obtained by combining the covered lines, branch outcomes, and killed mutants recorded for one to 25 suites per benchmark file.
Note that this analysis uses the individual suite measurements before structural merging and set-cover reduction.
We provide evidence later in Sect.~\ref{sec:rq3} that also after these steps all studied metrics significantly benefit from the combination. 

\begin{figure}[!htbp]
\centering
\includegraphics[width=\textwidth,height=0.72\textheight,keepaspectratio]{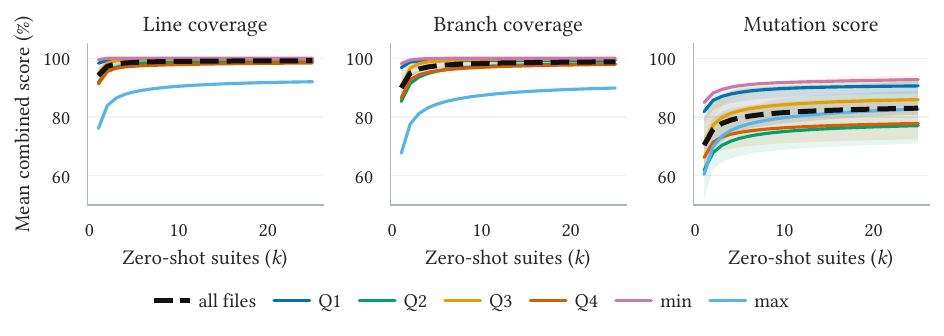}
\Description{Three line plots show mean combined line coverage, branch coverage, and mutation score as the number of zero-shot suites increases from one to 25.}
\caption{%
Expected scores from combining covered lines, branch outcomes, and killed mutants for $k$ of 25 zero-shot suites per file, averaged over three runs before merging and reduction.
Shaded bands show interquartile ranges overall and by cyclomatic complexity quartile (Q1 least, Q4 most complex).
Separate curves show the least and most complex files.%
}
\label{fig:phase1-score-accumulation-minmax-zoomed}
\end{figure}

Figure~\ref{fig:phase1-score-accumulation-minmax-zoomed} contains three plots showing the line and branch coverage as well as the mutation score over an increasing number of zero-shot suites $k$ for different complexity levels (complexity quartiles as well as the least and most complex file from the benchmark).  

We see that combining independently generated suites increases all three scores.
From one to 25 suites, mean line coverage rises from 94.1\% to 99.1\%, branch coverage from 89.9\% to 98.8\%, and mutation score from 70.4\% to 83.0\%.
Most of the increase occurs within the first five suites, followed by smaller gains as further suites are added.
We can see an improvement in every complexity quartile. For the most complex file, line coverage increases by over 15 percentage points, while branch coverage and mutation score each increase by over 20 percentage points.

\begin{rqbox}[title={Answer to RQ1: Parallelization of Zero-Shot Test Generation}]
Independent zero-shot suites complement one another, improving coverage and mutation score when combined.
The largest gains come from the first few suites, but also additional suites increase the coverage and continue to detect mutants that the smaller populations miss.
\end{rqbox}

\FloatBarrier
\subsection{RQ2: Comparison of Testuary with State-of-the-Art Test Generation Methods}
\label{sec:rq2}

For automated test generation, many effective methods have been proposed in the literature. Therefore, we analyze how Testuary compares with established test generators in line coverage, branch coverage, and mutation score. To answer this, we compare Testuary's final test suites with those produced by Pynguin, CodaMOSA, and CoverUp. Furthermore, we are also reporting the results for Zero-Shot as an additional baseline.

\begin{table}[!htbp]
\centering
\normalsize
\caption{Test-suite effectiveness and all-pairs comparisons. Means (\%) and sample SDs (standard deviation in percentage points) summarize file averages for three runs. Runs without passing suites score zero. Zero-Shot first averages 25 suites per run. Bold font marks the highest mean per metric. Superscripts identify significantly outperformed methods using paired file-level Wilcoxon signed-rank tests with Holm correction over 30 comparisons ($p<0.05$).}
\label{tab:rq2_scores_and_significance}
\setlength{\tabcolsep}{3pt}
\newlength{\scoreletterswidth}
\settowidth{\scoreletterswidth}{$^{abcd}$}
\begin{tabular*}{\textwidth}{@{\extracolsep{\fill}}lrrrrrr@{}}
\toprule
Method & \multicolumn{2}{c}{Line coverage} & \multicolumn{2}{c}{Branch coverage} & \multicolumn{2}{c}{Mutation score} \\
\cmidrule(lr){2-3}\cmidrule(lr){4-5}\cmidrule(lr){6-7}
 & Mean\hspace{\scoreletterswidth} & SD & Mean\hspace{\scoreletterswidth} & SD & Mean\hspace{\scoreletterswidth} & SD \\
\midrule
$^{a}$ Zero-Shot & 94.116\makebox[\scoreletterswidth][l]{$^{bc}$} & 6.232 & 89.914\makebox[\scoreletterswidth][l]{$^{bc}$} & 8.797 & 70.426\makebox[\scoreletterswidth][l]{$^{bc}$} & 13.921 \\
$^{b}$ Pynguin & 44.412\makebox[\scoreletterswidth][l]{$^{}$} & 25.552 & 31.238\makebox[\scoreletterswidth][l]{$^{}$} & 24.527 & 11.736\makebox[\scoreletterswidth][l]{$^{}$} & 11.177 \\
$^{c}$ CodaMOSA & 35.075\makebox[\scoreletterswidth][l]{$^{}$} & 32.106 & 23.846\makebox[\scoreletterswidth][l]{$^{}$} & 27.955 & 8.023\makebox[\scoreletterswidth][l]{$^{}$} & 9.985 \\
$^{d}$ CoverUp & 90.184\makebox[\scoreletterswidth][l]{$^{bc}$} & 14.998 & 87.552\makebox[\scoreletterswidth][l]{$^{bc}$} & 18.740 & 68.185\makebox[\scoreletterswidth][l]{$^{bc}$} & 20.957 \\
$^{e}$ Testuary (ours) & \textbf{99.995}\makebox[\scoreletterswidth][l]{$^{abcd}$} & 0.032 & \textbf{99.923}\makebox[\scoreletterswidth][l]{$^{abcd}$} & 0.343 & \textbf{98.640}\makebox[\scoreletterswidth][l]{$^{abcd}$} & 2.787 \\
\bottomrule
\end{tabular*}
\end{table}

When a generated test suite from a baseline approach fails to execute, we repeatedly remove tests that cannot be parsed or that fail, and evaluate the remaining suite once it passes. The baselines benefit from this way of evaluation. Runs without passing tests receive zero line coverage, branch coverage, and mutation scores.
For the statistical significance analyses, we use exact two-sided Wilcoxon signed-rank tests~\cite{wilcoxon1945individual} with Holm correction~\cite{holm1979simple} across all comparisons and all three metrics within each analysis.

\begin{figure}[!htbp]
\centering
\includegraphics[width=\textwidth,height=0.72\textheight,keepaspectratio]{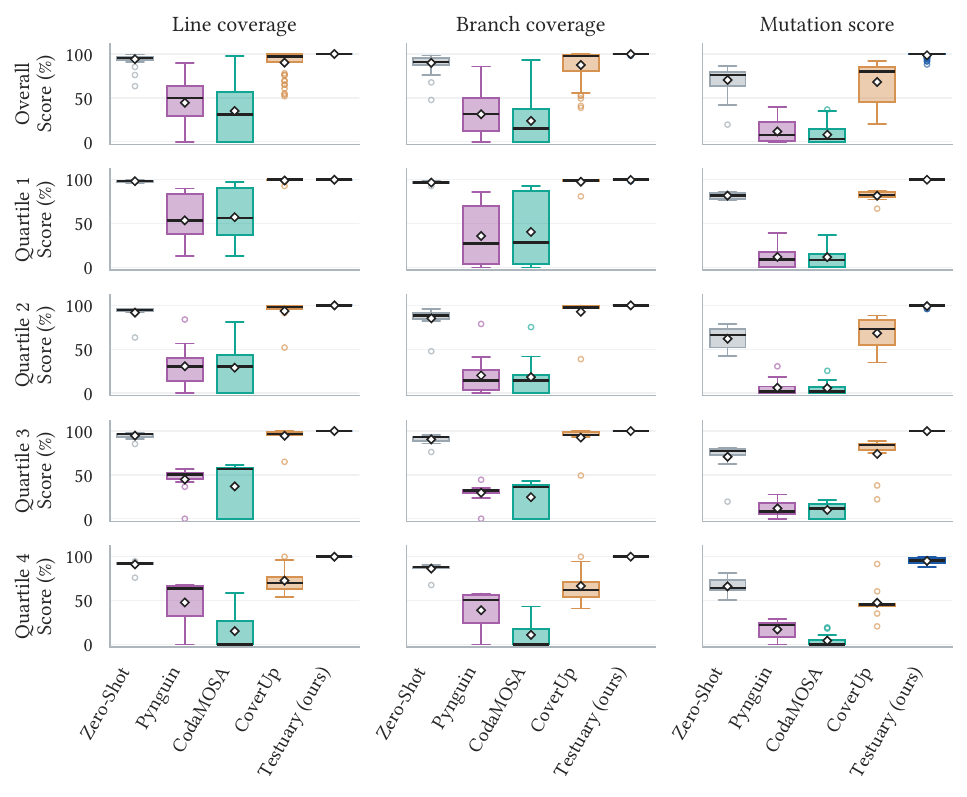}
\Description{Boxplot columns show line coverage, branch coverage, and mutation score by method. Rows show all files followed by the four complexity quartiles.}
\caption{%
Test suite effectiveness overall (top row) and by source-file cyclomatic complexity quartile (Q1 least, Q4 most complex).
Scores average three runs per file, including zero for runs without passing suites.
Zero-Shot first averages 25 suites per run.%
}
\label{fig:final-suite-adequacy-complexity}
\end{figure}

Table~\ref{tab:rq2_scores_and_significance} shows the means in percent and standard deviations (SD) for line and branch coverage as well as the mutation score for the studied approaches. Best results are written in bold font and superscripts identify significantly outperformed methods.

We see that Testuary achieves the highest mean scores for all three metrics, with line and branch coverage above 99.9\% and a mutation score of 98.64\%.
It significantly outperforms Pynguin, CodaMOSA, CoverUp, and Zero-Shot on each metric, with Holm-adjusted $p$-values below 0.001.
The mean mutation-score advantage is about 30 percentage points over CoverUp and about 28 percentage points over Zero-Shot.
Zero-Shot and CoverUp both significantly outperform Pynguin and CodaMOSA.

To analyze this in more detail, we present in Fig.~\ref{fig:final-suite-adequacy-complexity} box plots showing line and branch coverage as well as the mutation score for quartiles of the cyclomatic complexity (Q1 least, Q4 most complex) of the benchmark source file for all studied approaches. In addition, the top row shows box plots for the overall scores. 

We can see that Testuary achieves the highest mean scores in every quartile.
Its advantage over CoverUp is largest in Q4, where Testuary's mean line and branch coverage remain above 99.9\%, compared with 72.7\% and 66.5\% for CoverUp.
The mean mutation score gap grows from 18.3 percentage points in Q1 to 47.7 in Q4, where Testuary reaches 95.3\% and CoverUp 47.6\%.
CodaMOSA shows the largest Q1-to-Q4 drop in mean line coverage, from 57.3\% to 15.2\%, partly because it more often produces no passing suite in Q4.

CodaMOSA had no passing suite in 48 of 135 runs (35.6\%), compared with 19 runs (14.1\%) for Pynguin and one run (0.7\%) for CoverUp.
Testuary provides a passing suite in every run.
Note that runs without a passing suite count as zero for each metric, lowering the reported mean scores.

\begin{rqbox}[breakable=false,title={Answer to RQ2: Comparison of Testuary with State-of-the-Art Test Generation Methods}]
Testuary achieves significantly higher line coverage, branch coverage, and mutation scores than all four baseline methods in our experiment.
It achieves the highest mean scores across all four studied complexity quartiles.
\end{rqbox}

\FloatBarrier
\subsection{RQ3: Effectiveness Across Testuary's Phases}
\label{sec:rq3}

Testuary's test generation process consists of three phases that build on one another. It starts with a phase combining independently generated zero-shot test suites, followed by two agentic phases for improving coverage and mutation score. So the question is how the individual phases contribute to the overall result. To answer this question, we compare the individual phase outputs and also analyze the differences between the consecutive phases in detail. 

The handling of non-passing suites and the significance test procedure are the same as in Sect.~\ref{sec:rq2}.

\begin{figure}[!htbp]
\centering
\includegraphics[width=\textwidth,height=0.72\textheight,keepaspectratio]{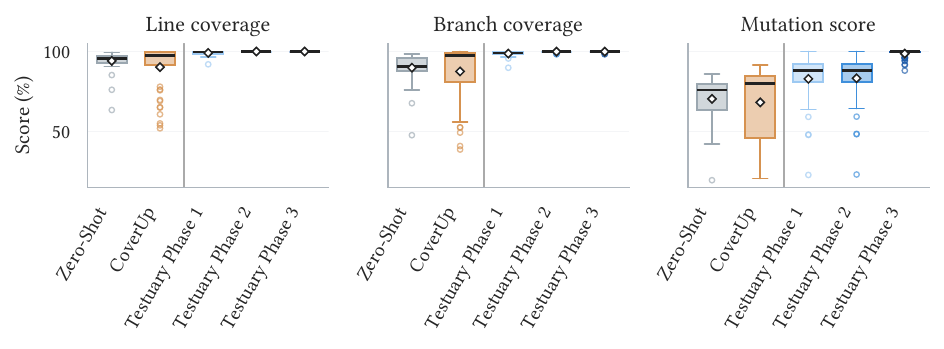}
\Description{Three box plot panels compare Zero-Shot, CoverUp, and the three Testuary phases for line coverage, branch coverage, and mutation score.}
\caption{%
Test suite effectiveness across Testuary's phases, with CoverUp and Zero-Shot for comparison.
Scores average three runs per file, including zero for runs without passing suites.
Zero-Shot first averages 25 suites per run.%
}
\label{fig:testuary-phase-progression-coverup}
\end{figure}

Figure~\ref{fig:testuary-phase-progression-coverup} shows box plots for the line and branch coverage as well as the mutation score for all three phases of Testuary. In addition to that, we added the results for Zero-Shot and CoverUp for a better comparison. Furthermore, Table~\ref{tab:rq3_paired_summary_with_zero_shot} presents the paired score differences from Zero-Shot through the three Testuary phases. Wins (W), ties (T), and losses (L) count positive, zero, and negative differences of the pairwise comparisons.

\begin{table}[!htbp]
\centering
\normalsize
\caption{Paired score differences from Zero-Shot through the three Testuary phases. Mean and median differences summarize paired file-level changes in percentage points. W (wins), T (ties), and L (losses) count positive, zero, and negative differences (first method minus second). Bold $p$-values indicate significance after Holm correction over nine tests ($p<0.05$).}
\label{tab:rq3_paired_summary_with_zero_shot}
\setlength{\tabcolsep}{3pt}
\renewcommand{\arraystretch}{1.1}
\begin{tabular*}{\textwidth}{@{\extracolsep{\fill}}lrrrr@{}}
\toprule
Metric & Mean difference & Median difference & W/T/L & Adjusted $p$ \\
\midrule
\multicolumn{5}{@{}l}{Testuary Phase~\circled{1} \textit{compared to} Zero-Shot} \\
Line coverage & 4.98 & 3.58 & 45/0/0 & $\boldsymbol{<0.001}$ \\
Branch coverage & 8.80 & 6.86 & 45/0/0 & $\boldsymbol{<0.001}$ \\
Mutation score & 12.46 & 12.11 & 45/0/0 & $\boldsymbol{<0.001}$ \\
\cmidrule[0.25pt](l{4pt}r{4pt}){1-5}
\multicolumn{5}{@{}l}{Testuary Phase~\circled{2} \textit{compared to} Testuary Phase~\circled{1}} \\
Line coverage & 0.90 & 0.44 & 27/18/0 & $\boldsymbol{<0.001}$ \\
Branch coverage & 1.20 & 1.04 & 30/15/0 & $\boldsymbol{<0.001}$ \\
Mutation score & 0.34 & 0.12 & 25/20/0 & $\boldsymbol{<0.001}$ \\
\cmidrule[0.25pt](l{4pt}r{4pt}){1-5}
\multicolumn{5}{@{}l}{Testuary Phase~\circled{3} \textit{compared to} Testuary Phase~\circled{2}} \\
Line coverage & $<0.01$ & 0.00 & 1/44/0 & $1.000$ \\
Branch coverage & $<0.01$ & 0.00 & 1/44/0 & $1.000$ \\
Mutation score & 15.41 & 10.73 & 43/2/0 & $\boldsymbol{<0.001}$ \\
\bottomrule
\end{tabular*}
\end{table}

The box plots show that Phase~\circled{1} achieves mean line coverage of 99.10\%, branch coverage of 98.72\%, and a mutation score of 82.89\%.
Compared with the average individual zero-shot suite, these are increases of 4.98, 8.80, and 12.46 percentage points, respectively (see Table~\ref{tab:rq3_paired_summary_with_zero_shot}).
All three scores improve for every file after averaging the three runs.

Phase~\circled{2} raises mean line and branch coverage above 99.9\%. As expected, for the mutation score we see only a slight increase of 0.34 percentage points.
Phase~\circled{3} then increases mutation score by 15.41 percentage points to 98.64\%, with almost no further change in coverage.
All improvements after Phases~\circled{1} and~\circled{2}, and the mutation-score improvement after Phase~\circled{3}, are statistically significant.
Furthermore, no file-level average score decreases between consecutive phases, showing that the improvements are kept across phases. 

Another important result is that Testuary's Phase~\circled{1} already significantly outperforms CoverUp in all three studied metrics.

\begin{rqbox}[title={Answer to RQ3: Effectiveness Across Testuary's Phases}]
Phase~\circled{1} provides high initial test coverage and improves mutation score through the combined zero-shot suites and already outperforms the baselines.
With Phase~\circled{2} it is possible to close most of the remaining coverage gaps.
Phase~\circled{3} contributes the largest mutation score increase, while maintaining the achieved coverage.
\end{rqbox}

\clearpage

\FloatBarrier
\subsection{RQ4: Model Usage and Test Suite Efficiency}
\label{sec:rq4}

\begin{figure}[!htbp]
\centering
\begin{subfigure}[t]{0.49\textwidth}
\centering
\captionsetup{position=top}
\caption{Model requests.}
\label{fig:model-response-use}
\includegraphics[width=\linewidth]{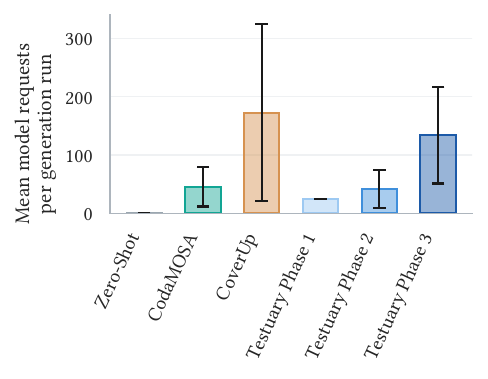}
\end{subfigure}\hfill
\begin{subfigure}[t]{0.49\textwidth}
\centering
\captionsetup{position=top}
\caption{Input and output tokens.}
\label{fig:model-token-use}
\includegraphics[width=\linewidth]{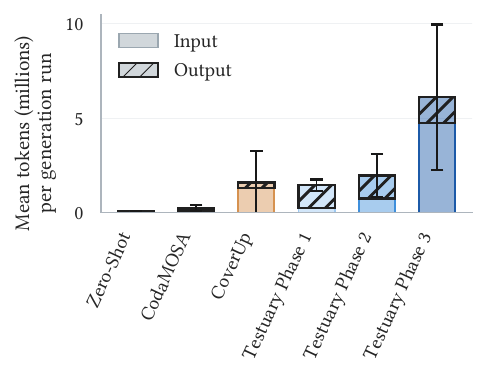}
\end{subfigure}
\caption{Model usage by method and Testuary phase.
Bars show means, with error bars indicating sample standard deviations across three-run file averages.
Testuary results are cumulative and report the counts measured after the individual phase.
Zero-Shot divides Phase~\protect\circled{1} totals by 25.
Requests count returned responses with recorded token usage.
Usage from runs without passing suites is kept.
Pynguin uses no LLM and is therefore excluded.}
\label{fig:model-use}
\Description{Side-by-side bar plots show model requests and token use, with input and hatched output tokens stacked for each method.}
\end{figure}

\begin{figure}[!htbp]
\centering
\begin{subfigure}[t]{0.49\textwidth}
\centering
\captionsetup{position=top}
\caption{Number of tests.}
\label{fig:number-of-tests}
\includegraphics[width=\linewidth]{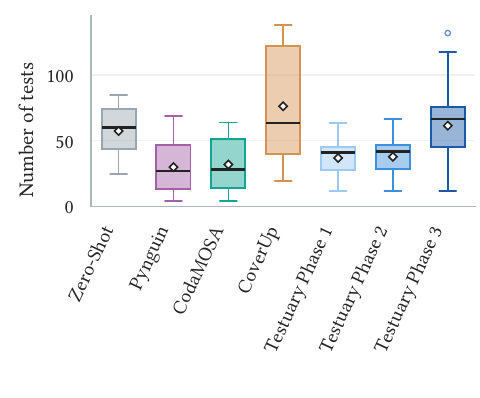}
\end{subfigure}\hfill
\begin{subfigure}[t]{0.49\textwidth}
\centering
\captionsetup{position=top}
\caption{Test suite runtime.}
\label{fig:test-suite-runtime}
\includegraphics[width=\linewidth]{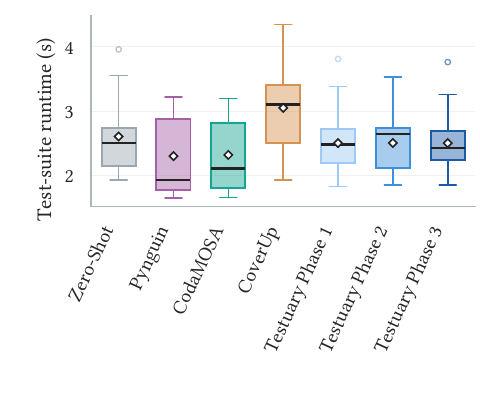}
\end{subfigure}
\caption{Test suite size and runtime on the same files, requiring three passing, timed runs per method.
Counts and runtimes are averaged per file, with Zero-Shot first averaging its 25 suites per run.
Non-passing suites are excluded and not assigned zero.}
\label{fig:delivered-suite-size-common}
\Description{Side-by-side boxplots compare test counts and runtimes on identical files with three passing runs for every method.}
\end{figure}

\begin{figure}[!htbp]
\centering
\includegraphics[width=\textwidth,height=0.72\textheight,keepaspectratio]{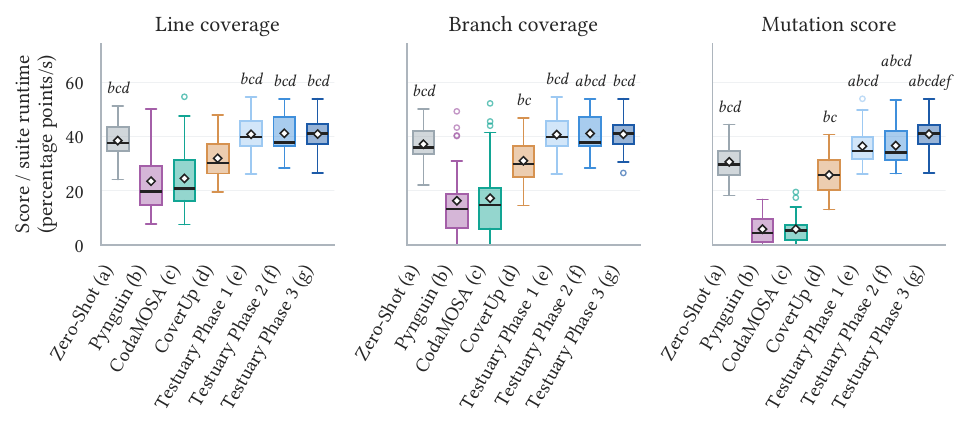}
\Description{Three boxplot panels compare line coverage, branch coverage, and mutation score per second of test suite runtime across methods and Testuary phases.}
\caption{%
Test suite effectiveness per second.
Each file's three-run mean score is divided by its mean test suite runtime.
All methods use the same files, requiring three passing, timed runs each.
Zero-Shot first averages 25 suite scores and runtime medians per run.
Missing runtimes are excluded.
Letters above boxes identify significantly outperformed methods, using paired Wilcoxon signed-rank tests with Holm correction ($p<0.05$).%
}
\label{fig:effectiveness-per-suite-second-common}
\end{figure}

Integrating LLMs into test generation methods makes the number of API requests and token usage become relevant metrics. Furthermore, for the practical deployment of the generated test suites, it is also relevant how effective they are relative to their execution time. Therefore, to answer RQ4, we analyze for each of Testuary's phases as well as the baseline approaches the number of model requests (Fig.~\ref{fig:model-response-use}), the input and output tokens (Fig.~\ref{fig:model-token-use}), the number of tests in a suite (Fig.~\ref{fig:number-of-tests}), the overall test suite runtime in seconds (Fig.~\ref{fig:test-suite-runtime}), and the achieved scores relative to the test suite's runtime (Fig.~\ref{fig:effectiveness-per-suite-second-common}). 

Completing all three Testuary phases requires 133.9 model requests on average, compared with 172.9 for CoverUp and 45.7 for CodaMOSA (Fig.~\ref{fig:model-response-use}).
However, Testuary uses more tokens, averaging 6.11 million input and output tokens per generation run, compared with 1.59 million for CoverUp and 0.24 million for CodaMOSA (Fig.~\ref{fig:model-token-use}).
Within Testuary, cumulative token use rises from 1.46 million after Phase~\circled{1}, which already beats CoverUp, to 1.97 million after Phase~\circled{2}, and 6.11 million after Phase~\circled{3}. 
Most output tokens are used during Testuary's initial generation, while most input tokens are used in the agentic phases.

Among the suites passing, Testuary's final suites contain 61.4 tests and execute in 2.50 seconds on average, compared with 76.3 tests and 3.05 seconds for CoverUp (Fig.~\ref{fig:delivered-suite-size-common}).
Pynguin and CodaMOSA produce smaller suites with lower mean runtimes.
Across Testuary's phases, mean runtime remains close to 2.50 seconds even as the mean number of tests increases from 36.8 after Phase~\circled{1} to 61.4 after Phase~\circled{3}.

Considering effectiveness relative to runtime, Testuary's phases achieve significantly higher scores per second than Pynguin, CodaMOSA, and CoverUp for all three metrics (Fig.~\ref{fig:effectiveness-per-suite-second-common}).
For mutation score per second, Phase~\circled{3} also significantly outperforms Zero-Shot and both preceding phases. 

\begin{rqbox}[breakable=false,title={Answer to RQ4: Model Usage and Test Suite Efficiency}]
Testuary uses fewer model requests than CoverUp but uses more tokens.
Testuary's final suites contain fewer tests and have a lower mean runtime than CoverUp's.
Furthermore, Testuary's phases achieve significantly higher scores relative to the test suite’s
runtime than Pynguin, CodaMOSA, and CoverUp for the studied metrics.
Testuary's mutation score per second in Phase~\circled{3} also significantly outperforms Zero-Shot and both of its preceding phases.
\end{rqbox}

\section{Threats to Validity}\label{sec:limitations}

The outputs of LLMs are nondeterministic, and also the search-based baselines rely on random decisions, which may influence the results. To reduce this risk, we run each method three times, totaling \$1,102.45 in API costs (for development and experimental runs). 
Across the three repetitions, Testuary's overall scores have a low variance, indicating stable results.

As the benchmark code and the existing project tests are publicly available, they could have been part of the model's training data. However, this risk exists in any evaluation with public benchmarks and LLMs. To mitigate this risk, all LLM-based methods use the same model, so any prior knowledge is available to all methods and does not explain the observed differences. 

Line and branch coverage only show which code is executed by the tests, not whether the tests detect faults. To reduce this risk, we additionally measure the mutation score, for which prior work shows a relation between the detection of mutants and the detection of real faults~\cite{just2014mutants}. This combination of coverage and mutation score matches the standard for evaluating test generation methods and is also used in TestGenEval~\cite{jain2025testgeneval}. 

The validity of the mutation score assumes that tests detect mutants through the program's behavior. During development we occasionally observed generated tests that instead compared the source code or its hash against a fixed value. To avoid such tests, Testuary applies the checks described in Sects.~\ref{sec:phase1} and~\ref{sec:agent-loops} in all phases, which detect and remove such tests.

\section{Conclusions and Future Work}\label{sec:conclusion}

Agentic test generation methods improve a given test suite step by step, so a weak starting point may lead to many iterations. Therefore, we presented in this paper Testuary, which gives the agentic phases a head start. It merges independently generated zero-shot test suites into a single suite and then minimizes it using a runtime-weighted set-cover approach. After that, two agentic phases target the remaining gaps in coverage and mutation score.

We compared Testuary against Zero-Shot, Pynguin, CodaMOSA, and CoverUp. We observed that the combined zero-shot suites improve all three metrics over the average single zero-shot suite, with most of the gain being achieved by combining the first few suites. Already after this non-agentic phase, Testuary significantly outperforms CoverUp in line coverage, branch coverage, and mutation score. After the agentic phases, Testuary reaches on average more than 99.9\% line and branch coverage along with a mutation score of 98.6\% (CoverUp 68.2\%). It also achieves the highest scores even in the quartile of the most complex benchmark files. Testuary needs fewer model requests than CoverUp but uses more tokens, especially in the mutation phase. The resulting suites are on average smaller and faster than CoverUp's and achieve significantly higher scores per second of test runtime than Pynguin, CodaMOSA, and CoverUp.

In future work, we plan to extend Testuary to support additional programming languages and to reduce the token consumption of the agentic phases.

\bibliographystyle{ACM-Reference-Format}
\bibliography{sample-base}

\clearpage
\appendix
\section{Appendix: Additional Evaluation Results}
\label{sec:additional-results}

In the following, we provide supplementary analyses of the presented results. Appendix~\ref{sec:app_detected_mutants} compares which mutants are detected by each technique, thereby complementing the mutation-score comparison for RQ2 in Section~\ref{sec:rq2} with an analysis of the overlap between the detected mutants. Appendix~\ref{sec:app_variation} reports the mean scores for each approach in each run, complementing the run-averaged results of Table~\ref{tab:rq2_scores_and_significance} (also RQ2) as well as the discussion of result stability in Section~\ref{sec:limitations}.

\subsection{Overlap in Detected Mutants}
\label{sec:app_detected_mutants}

\begin{figure}[!htbp]
\centering
\includegraphics[width=0.84\textwidth,height=0.72\textheight,keepaspectratio]{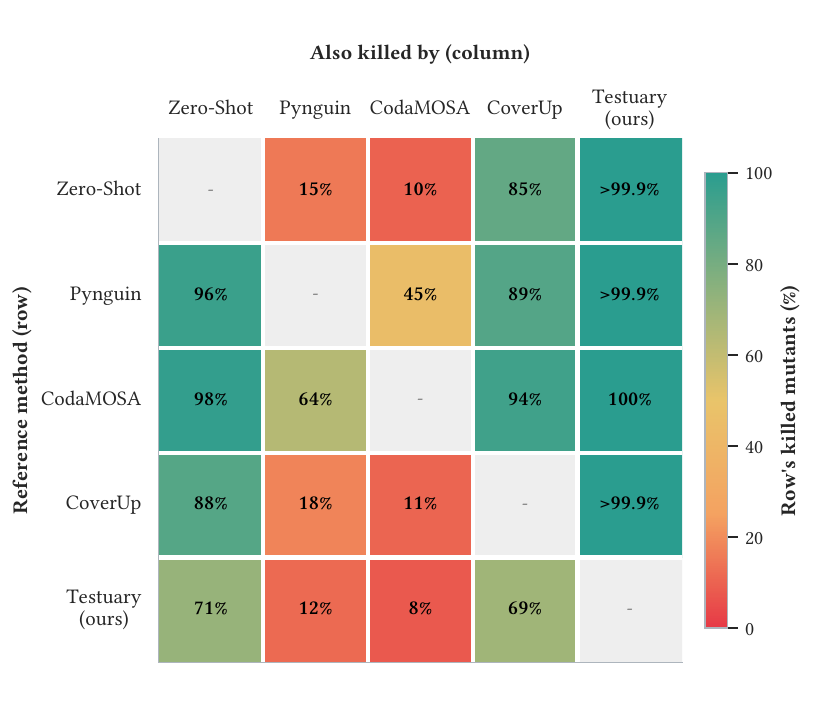}
\Description{A directional matrix shows the percentage of each row method's detected mutants also detected by each column method.}
\caption{%
Percentage of the row method's killed mutants also killed by the column method, averaged within each file and then across files.
Zero-Shot represents individual suites, not their union.
Within-method comparisons are not shown.%
}
\label{fig:appendix-mutation-overlap}
\end{figure}

Figure~\ref{fig:appendix-mutation-overlap} compares the mutants detected by the methods rather than only their mutation scores.
Each cell gives the percentage of mutants killed by the row method that are also killed by the column method.
Each of the three runs of the row method is compared with each of the three runs of the column method.
For each file, we average the resulting nine fractions and then average across files.
For Zero-Shot, each run contributes its 25 individual suites, whose fractions are averaged within each pair of runs.
An empty reference kill set has no defined fraction and is omitted, while an empty comparison kill set contributes zero.
The rows of Pynguin and CodaMOSA are therefore based on 77.8\% and 57.8\% of the files.
On average, Testuary detects more than 99.9\% of the mutants detected by each baseline in this comparison.
The reverse fractions are lower because Testuary detects additional mutants.
CoverUp detects most of the mutants detected by Pynguin and CodaMOSA, while these detect only a small share of those detected by CoverUp.

\clearpage
\subsection{Variation Across Runs}
\label{sec:app_variation}

Table~\ref{tab:appendix-runs} shows the mean scores of each method over all runs and separately for each of the three runs, as well as the standard deviation of the three run means.
As in the main analysis, runs without a passing suite score zero.
For every metric, the methods rank in the same order in each run.
Testuary's run means are nearly identical.
This low variance is also a result of Testuary often reaching the maximum score of 100\%, which leaves limited room for variation.
CoverUp varies most for all three metrics.

\begin{table}[!htbp]
\centering
\normalsize
\caption{Mean score of each method over all runs and in each run, across all files, and the sample standard deviation (SD) of the three run means. Parentheses give the difference between a run's mean and the mean over all runs, in percentage points, in green when the run is higher and in red when it is lower. Runs without a passing suite score zero, and Zero-Shot averages its 25 suites per run.}
\label{tab:appendix-runs}
\setlength{\tabcolsep}{3pt}
\renewcommand{\arraystretch}{1.6}
\begin{tabular*}{\textwidth}{@{\extracolsep{\fill}}lrr@{\extracolsep{0pt}\hspace{2pt}}l@{\extracolsep{\fill}\hspace{6pt}}r@{\extracolsep{0pt}\hspace{2pt}}l@{\extracolsep{\fill}\hspace{6pt}}r@{\extracolsep{0pt}\hspace{2pt}}l@{\extracolsep{\fill}\hspace{6pt}}c@{}}
\toprule
 & \multicolumn{7}{c}{Mean score in \% (difference to all runs in parentheses)} & \\
\cmidrule(lr){2-8}
Method & All runs & \multicolumn{2}{c}{Run 0} & \multicolumn{2}{c}{Run 1} & \multicolumn{2}{c}{Run 2} & \smash{\renewcommand{\arraystretch}{1}\begin{tabular}[b]{@{}c@{}}SD of\\ run means\end{tabular}} \\
\midrule
\multicolumn{9}{@{}l}{\textit{Line coverage}} \\
Zero-Shot & 94.116 & 93.913 & \textcolor{red!75!black}{($-$0.203)} & 94.285 & \textcolor{green!45!black}{($+$0.169)} & 94.149 & \textcolor{green!45!black}{($+$0.033)} & 0.188 \\
Pynguin & 44.412 & 43.613 & \textcolor{red!75!black}{($-$0.799)} & 43.581 & \textcolor{red!75!black}{($-$0.831)} & 46.041 & \textcolor{green!45!black}{($+$1.629)} & 1.411 \\
CodaMOSA & 35.075 & 34.981 & \textcolor{red!75!black}{($-$0.094)} & 34.900 & \textcolor{red!75!black}{($-$0.175)} & 35.343 & \textcolor{green!45!black}{($+$0.268)} & 0.236 \\
CoverUp & 90.184 & 88.197 & \textcolor{red!75!black}{($-$1.987)} & 92.111 & \textcolor{green!45!black}{($+$1.927)} & 90.243 & \textcolor{green!45!black}{($+$0.059)} & 1.957 \\
Testuary (ours) & \textbf{99.995} & \textbf{100.000} & \textcolor{green!45!black}{($+$0.005)} & \textbf{99.986} & \textcolor{red!75!black}{($-$0.009)} & \textbf{100.000} & \textcolor{green!45!black}{($+$0.005)} & 0.008 \\
\cmidrule[0.25pt](l{4pt}r{4pt}){1-9}
\multicolumn{9}{@{}l}{\textit{Branch coverage}} \\
Zero-Shot & 89.914 & 89.655 & \textcolor{red!75!black}{($-$0.259)} & 90.093 & \textcolor{green!45!black}{($+$0.179)} & 89.994 & \textcolor{green!45!black}{($+$0.080)} & 0.230 \\
Pynguin & 31.238 & 30.920 & \textcolor{red!75!black}{($-$0.318)} & 30.475 & \textcolor{red!75!black}{($-$0.763)} & 32.320 & \textcolor{green!45!black}{($+$1.082)} & 0.963 \\
CodaMOSA & 23.846 & 23.797 & \textcolor{red!75!black}{($-$0.049)} & 23.578 & \textcolor{red!75!black}{($-$0.268)} & 24.163 & \textcolor{green!45!black}{($+$0.317)} & 0.295 \\
CoverUp & 87.552 & 84.712 & \textcolor{red!75!black}{($-$2.840)} & 90.018 & \textcolor{green!45!black}{($+$2.466)} & 87.926 & \textcolor{green!45!black}{($+$0.374)} & 2.673 \\
Testuary (ours) & \textbf{99.923} & \textbf{99.927} & \textcolor{green!45!black}{($+$0.004)} & \textbf{99.916} & \textcolor{red!75!black}{($-$0.007)} & \textbf{99.927} & \textcolor{green!45!black}{($+$0.004)} & 0.006 \\
\cmidrule[0.25pt](l{4pt}r{4pt}){1-9}
\multicolumn{9}{@{}l}{\textit{Mutation score}} \\
Zero-Shot & 70.426 & 70.141 & \textcolor{red!75!black}{($-$0.285)} & 70.749 & \textcolor{green!45!black}{($+$0.323)} & 70.390 & \textcolor{red!75!black}{($-$0.036)} & 0.306 \\
Pynguin & 11.736 & 11.653 & \textcolor{red!75!black}{($-$0.083)} & 11.574 & \textcolor{red!75!black}{($-$0.162)} & 11.981 & \textcolor{green!45!black}{($+$0.245)} & 0.216 \\
CodaMOSA & 8.023 & 7.569 & \textcolor{red!75!black}{($-$0.454)} & 8.324 & \textcolor{green!45!black}{($+$0.301)} & 8.177 & \textcolor{green!45!black}{($+$0.154)} & 0.400 \\
CoverUp & 68.185 & 65.852 & \textcolor{red!75!black}{($-$2.333)} & 69.723 & \textcolor{green!45!black}{($+$1.538)} & 68.980 & \textcolor{green!45!black}{($+$0.795)} & 2.054 \\
Testuary (ours) & \textbf{98.640} & \textbf{98.656} & \textcolor{green!45!black}{($+$0.016)} & \textbf{98.633} & \textcolor{red!75!black}{($-$0.007)} & \textbf{98.631} & \textcolor{red!75!black}{($-$0.009)} & 0.014 \\
\bottomrule
\end{tabular*}
\end{table}

\end{document}